\documentclass[conference]{IEEEtranSPMB}

\IEEEoverridecommandlockouts
\ifCLASSINFOpdf
\else
\fi
\usepackage{amsmath,graphicx,soul,subcaption,balance}
\usepackage{balance}
\usepackage[numbers,sort&compress]{natbib}

\usepackage{parskip}
\usepackage{float} 

\usepackage{mathptmx}

\usepackage{multirow} 

\usepackage{color}
\usepackage{soul,mathrsfs}
\usepackage[utf8]{inputenc}
\usepackage[T1]{fontenc}
\usepackage{array}
\usepackage[hyphens, spaces, obeyspaces]{url}

\usepackage[papersize={8.5in,11in}, left=1in, right=1in, top=1in, bottom=1in]{geometry}
\newcolumntype{L}[1]{>{\raggedright\let\newline\\\arraybackslash\hspace{0pt}}m{#1}}
\newcolumntype{C}[1]{>{\centering\let\newline\\\arraybackslash\hspace{0pt}}m{#1}}
\newcolumntype{R}[1]{>{\raggedleft\let\newline\\\arraybackslash\hspace{0pt}}m{#1}}

\usepackage[singlelinecheck=false,justification=centering]{caption}
\DeclareCaptionLabelFormat{mylabel}{#1 #2.\hspace{0.7ex}}
\renewcommand{\thetable}{\arabic{table}}

\usepackage{xcolor,colortbl}
\definecolor{light-gray}{gray}{0.83}

\newcommand{\spmbtitlefont}{\fontsize{11.0pt}{11.00pt}\selectfont\bf\vspace{0.7em}}
\newcommand{\spmbauthorfont}{\fontsize{11.0pt}{11.0pt}\selectfont\vspace{0em}}
\newcommand{\subparagraph}{}
\usepackage[compact]{titlesec}
\titleformat{\section}
   {\center\normalfont\sc}{\thesection.}{0.7ex}{}
\titlespacing{\section}{0pt}{2ex}{1.5ex}
\titlespacing{\subsection}{0pt}{1.5ex}{1.2ex}
\titlespacing{\subsubsection}{0pt}{1ex}{0.9ex}
\makeatletter
\renewcommand*{\@seccntformat}[1]{\csname the#1\endcsname .\hspace{0.7em}}
\makeatother

\usepackage{fancyhdr,lastpage}
\fancypagestyle{firststyle}
{
   \fancyhf{}
	\lfoot{ }\rfoot{ }
}
\title{\spmbtitlefont Dilated Inception U-Net (DIU-Net) for Brain Tumor Segmentation
{\vspace{-2.4\baselineskip}
}
}
    \author{\spmbauthorfont\IEEEauthorblockN{
    D. Cahall\textsuperscript{\it 1}, 
    G. Rasool\textsuperscript{\it 1}, 
    N. C. Bouaynaya\textsuperscript{\it 1}, and
    H. M. Fathallah-Shaykh\textsuperscript{\it 2}
    }
    \vspace{0.9em}
    \IEEEauthorblockA{\spmbauthorfont
        1. Department of Electrical and Computer Engineering, Rowan University, New Jersey, USA \\
        2. Departments of Neurology and Mathematics, University of Alabama at Birmingham, USA \\
        danielenricocahall@gmail.com, rasool@rowan.edu, bouaynaya@rowan.edu, hfshaykh@uabmc.edu
    }
}

\newcommand{\AbstractSummary}{D.\ Cahall, et al.: Dilated Inception U-Net (DIU-Net) for Brain Tumor Segmentation}

\begin{document}

\IEEEaftertitletext{}
\maketitle

\begin{abstract}

Magnetic resonance imaging (MRI) is routinely used for brain tumor diagnosis, treatment planning, and post-treatment surveillance. Recently, various models based on deep neural networks have been proposed for the pixel-level segmentation of tumors in brain MRIs. However, the structural variations, spatial dissimilarities, and intensity inhomogeneity in MRIs make segmentation a challenging task. We propose a new end-to-end brain tumor segmentation architecture based on U-Net that integrates Inception modules and dilated convolutions into its contracting and expanding paths. This allows us to extract local structural as well as global contextual information. We performed segmentation of glioma sub-regions, including tumor core, enhancing tumor, and whole tumor using Brain Tumor Segmentation (BraTS) 2018 dataset. Our proposed model performed significantly better than state-of-the-art U-Net-based model ($p<0.05$) for tumor core and whole tumor segmentation.
\end{abstract}

\IEEEpeerreviewmaketitle    
\thispagestyle{firststyle}  
\section{Introduction}
\label{sec:intro}
Machine learning techniques based on deep neural networks have become increasingly common in the medical imaging field in recent years ~\cite{litjens_kooi_bejnordi_setio_ciompi_ghafoorian_laak_ginneken_sanchez_2017}. One of the challenging problems in medical imaging is the pixel level segmentation of various biological structures in a given image, e.g., segmentation of brain tumors in MRIs \cite{Fathallah-Shaykh2018, cahall2019inception}. Accurate and timely segmentation of brain tumors can help physicians with the diagnosis, treatment planning, and post-treatment surveillance \cite{Fathallah-Shaykh2018}.

The accurate segmentation of various structures in an image is dependent upon the extraction of local structural and global contextual information. Several multi-path architectures have been proposed in the medical image segmentation literature which extract information from given data at multiple scales \cite{salehi_erdogmus_gholipour_2017, havaei2017brain, kamnitsas_ledig_newcombe_simpson_kane_menon_rueckert_glocker_2017}. U-Net, proposed by Ronneberger et al., is commonly used for the segmentation of various structures in medical images \cite{ronneberger_fischer_brox_2015}. U-Net is built using (1) a contracting path, which captures high-resolution, contextual features while downsampling at each layer, and (2) an expanding path, which increases the resolution of the output through upsampling at each layer \cite{ronneberger_fischer_brox_2015}. The features from the contracting path are fused with features from the expanding path through long skip connections, ensuring localization of the extracted contextual features \cite{drozdzal_vorontsov_chartrand_kadoury_pal_2016}. U-Net was originally developed and applied to cell tracking; however, more recently, the model has been applied to other medical segmentation tasks, such as brain vessel segmentation, brain tumor segmentation, and retinal segmentation \cite{livne_rieger_aydin_taha_akay_kossen_sobesky_kelleher_hildebrand_frey_et_al, dong_yang_liu_mo_guo_2017, girard2019joint}. Variations of U-Net, such as 3D U-Net, GRA U-Net, RIC-UNet, PsLSNet, and SDResU-Net, have been proposed to tackle different segmentation problems in medical imaging \cite{kamnitsas_ledig_newcombe_simpson_kane_menon_rueckert_glocker_2017, dash2019pslsnet, sdresunet, zeng_xie_zhang_lu_2019}.  
 
The concept of extracting and aggregating features at multiple scales has also been accomplished by Inception modules \cite{szegedy_liu_jia_sermanet_reed_anguelov_erhan_vanhoucke_rabinovich_2015}. However, the mechanism of multi-scale feature extraction is different compared to multi-path architectures \cite{salehi_erdogmus_gholipour_2017, havaei2017brain, kamnitsas_ledig_newcombe_simpson_kane_menon_rueckert_glocker_2017}. Each Inception module applies filters of various sizes at each layer and concatenates resulting feature maps \cite{szegedy_liu_jia_sermanet_reed_anguelov_erhan_vanhoucke_rabinovich_2015}. Inception modules within U-Net have also been recently proposed for brain tumor segmentation, left atrial segmentation, and liver segmentation \cite{li2019novel, cahall2019inception, wang2019ensemble, song2020bottleneck}. 

Several extensions and modifications to the Inception module have been proposed, such as dilated (also known as atrous) convolutions \cite{yu2015multi}. Dilated convolutions enable the learned filters in a convolutional neural network (CNN) to have larger receptive fields with fewer parameters, thereby reducing the computational cost. Inception modules using dilated convolutions have also been utilized to improve image resolution, visual saliency prediction, change detection in multi-sensor images, and learning optical flow \cite{shi2017single, yang2019predicting, wang2020deep, zhai2019learning}. Recently introduced dilated residual Inception block accomplish multi-scale feature extraction in an end-to-end, fully convolutional retinal depth estimation model \cite{shankaranarayana_ram_mitra_sivaprakasam_2019}. 

We introduce an end-to-end brain tumor segmentation framework based on U-Net architecture with dilated Inception modules, referred to as Dilated Inception U-Net (DIU-Net), to accomplish multi-scale feature extraction. We demonstrate that integrating dilated convolutions within Inception modules results in significant improvement ($p<0.05$) in the segmentation of two of the three glioma sub-regions, i.e., tumor core and whole tumor.

\section{Methods} \label{sec:methods}

\subsection{Dilated Inception U-Net (DIU-Net) Architecture}
\begin{figure*}[htpb]
\centering
\includegraphics[height=3in,width=6.5in]{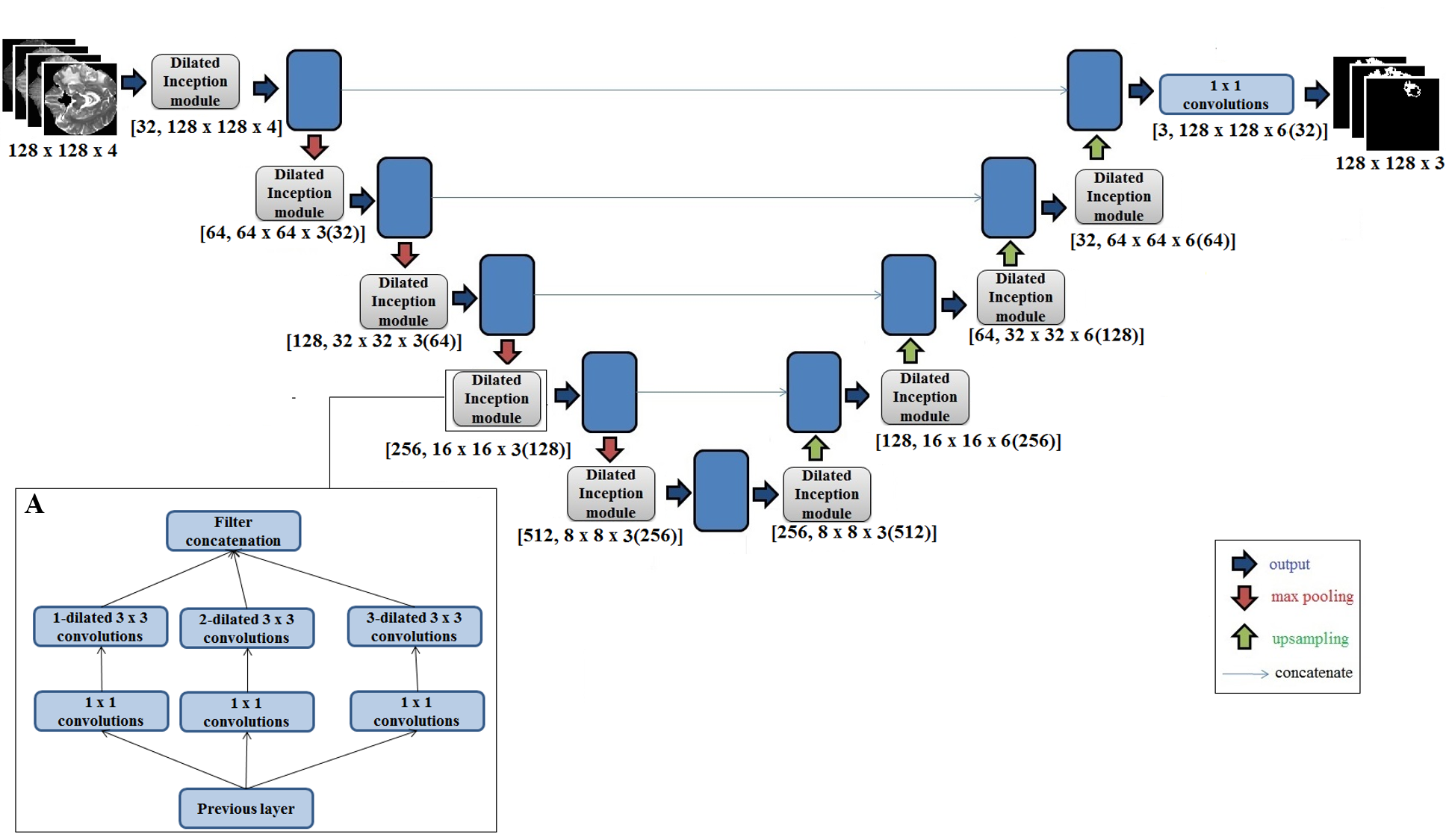}
\caption{\label{fig:arch} DIU-Net architecture with contracting and expanding path and a bottleneck in the middle is presented. The set of numbers shown below each Inception module indicates the total number of filters used, height, width, and depth of the input feature map. On the contracting path, the multiplication by 3 indicates three $l$-dilated convolutional filters. On expanding path, the concatenation of the feature maps from the contracting path doubles the depth of the output feature map, hence the multiplication by 6. (A) Dilated Inception module with three $l$-dilated convolutional filters and $1 \times 1$ dimensional reduction convolution filters is presented. }
\end{figure*}

We propose to integrate dilated convolutions and Inception modules in the U-Net architecture \cite{shi2017single}. In our settings, each dilated Inception module consists of three $1 \times 1$ convolution operations, each followed by one $l$-dilated convolutional filter with $l = $1, 2, and 3. The $1 \times 1$ convolution filters perform dimensionality reduction, while three $l$-dilated convolutional filters each of size $3 \times 3 $  implement atrous convolutions. The schematic layout of a dilated inception module is provided in Fig. \ref{fig:arch}A and a detailed description of dilated convolution filters is provided in Section \ref{sec:Dilted_Conv}. Finally, we use the rectified linear unit (ReLU) as the activation function and performed batch normalization in each dilated Inception module \cite{ioffe2015batch}. 

In Fig. \ref{fig:arch}, we present the detailed architecture of our proposed DIU-Net. We used a contracting-expanding architecture, resembling U-Net, with a bottleneck in the middle. The number of filters double at each layer on the contracting side and halve on the expanding side. On the other hand, the size of the output feature map (height and width) halves on the contracting side and doubles on the expanding side. We perform downsampling using max-pooling on the contracting path and upsampling on the expanding path. We also perform feature concatenation on the expanding path, i.e., features from the corresponding layer of the contracting path are concatenated with those on the expanding path. At the last layer on the expanding path, the output height and width are equal to the height and width of the original input images. At the output, we perform 1 $\times$ 1 convolutions to reduce the depth of the last feature map equal to the number of segmentation classes (i.e., tumor regions). Finally, a pixel-wise activation is performed to convert feature maps into binary segmentation outputs.

\subsection{Dilated Convolutions} \label{sec:Dilted_Conv}
We consider an image $I$ of size $m \times n$ and a discrete convolutional filter $w$ of size $k \times k$. The linear convolutional operation between the image $I$ and the filter $w$ is given by:
\begin{equation} \label{eq:conv}
(I*w)(p) = \sum_{s} I[p + s]w[s].
\end{equation}

The simple convolution operation can be generalized to $l$-dilated convolution ($*_l$) as \cite{yu2015multi}:

\begin{equation} \label{eq:dil}
(I*_lw)(p) = \sum_{s} I[p + ls]w[s].
\end{equation} 

It is evident that for $l=1$, we get the the simple convolutional operation given in \ref{eq:conv}. However, for $l > 1$, $l-1$ zeroes are inserted between each filter element, creating a $k_s \times k_s$ scaled and sparse filter, where $k_s$ is defined as:
\begin{align}
    k_s &= k + (k - 1)(l-1), \\
    &= l(k-1) + 1.
\end{align}

The scaling $s$ increases the receptive field of the filter by a factor $\left(\frac{k_s}{k}\right)^2$.

\begin{align}
    \frac{k_s}{k} &= \frac{k + (k - 1)(l-1)}{k}, \\
    &= l + \left(\frac{-l + 1}{k}\right). 
\end{align}

\begin{figure}[tb]
\centering
\includegraphics[width=3.3in]{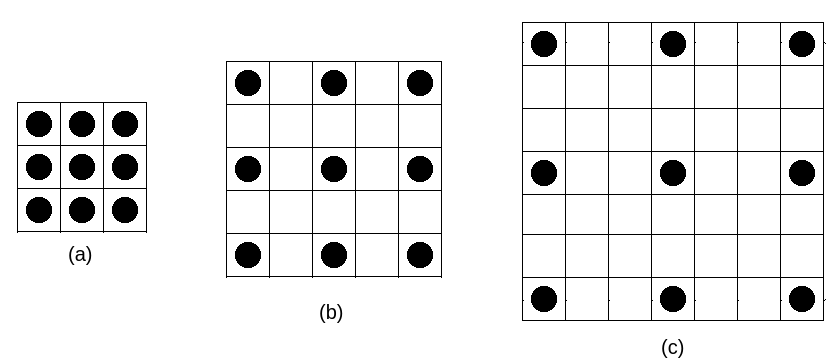} 
\caption{\label{fig:dil_conv} Three cases of a $3 \times 3$ dilated filter with $l=1, 2, \text{ and } 3$, are presented in sub-figures (a), (b), and (c) respectively. We note that the number of filter elements, indicated by the black dots, stays constant while the receptive field increases proportionally to $l$.  }
\end{figure}

The receptive field of the filter increases linearly with $l$, while the number of elements ($k \times k$) remains fixed. In Fig. \ref{fig:dil_conv}, we present $l$-dilated convolution filters of size $3 \times 3$ for $l=1, 2, $ and 3. 

\subsection{Dataset and Pre-Processing}
We used BRATS 2018 dataset for our experiments \cite{bakas2018identifying}. The dataset includes MRIs of 210 high-grade glioma (HGG) and 75 low-grade glioma (LGG) patients. Each patient's data consists of four MRI sequences: T2-weighted (T2), T1, T1 with gadolinium enhancing contrast (T1C), and Fluid-Attenuated Inversion Recovery (FLAIR) images. BRATS also provides pixel-level manual segmentation markings for three \emph{intra-tumoral} structures:  necrotic and non-enhancing tumor core (label = 1), peritumoral edema (label = 2), and enhancing tumor (label = 4). From the intra-tumoral structures, the following \emph{glioma sub-regions} \cite{menze:hal-00935640} were defined: whole tumor (WT) which encompasses all three intra-tumoral structures (i.e., label =  $1 \cup 2 \cup 4$), tumor core (TC) that contains all but the peritumoral edema (i.e., label = $1 \cup 4 $), and enhancing tumor (ET) (label = $ 4 $), where $\cup$ represents union operation.

The BRATS dataset is provided in a preprocessed format, i.e., all the images are skull-stripped, resampled to an isotropic 1mm$^3$ resolution, and all four modalities of each patient are co-registered. We applied additional pre-processing that included (in order): 1) computing the bounding box of the brain in each image, and extracting the selected portion of the image, effectively zooming in on the brain and discounting excess background pixels, 2) re-sizing the cropped image to $128 \times 128$ pixels, 3) discarding images which contained no tumor regions in the ground truth segmentation, 4) applying an intensity windowing function to each image such that the lowest 1\% and highest 99\% of pixel values were mapped to 0 and 255, respectively, and 5) applying z-score normalization to each image i.e., subtracting the mean and dividing by the standard deviation of the dataset. 

The input to DIU-Net is an $N \times M \times D$ pixel image, where $ N=M=128 $ pixels and $ D = 4 $ which represents four MRI modalities. The output of the model is an $N \times M \times K$ tensor, where $ K = 3 $ and represents total number of segmentation classes, i.e., three intra-tumoral structures. Each slice of $K$ is a binary image and represents the predicted segmentation for the $i$\textsuperscript{th} class where 0 $\leq i \leq K-1$.

\subsection{Evaluation Metric and the Loss Function }
Dice Similarity Coefficient or simply the Dice score is extensively used for the evaluation of segmentation algorithms in medical imaging applications \cite{bakas_akbari_sotiras_bilello_rozycki_kirby_freymann_farahani_davatzikos_2017}. The Dice scores between a predicted binary image \textit{P} and a ground truth binary image \textit{G}, both of size \textit{N} $\times$ \textit{M} is given by:

\begin{equation} \label{eq:1}
\text{Dice}(P, G) = {\frac{2\sum_{i=0}^{N-1}\sum_{j=0}^{M-1}P_{ij} G_{ij}} {\sum_{i=0}^{N-1}\sum_{j=0}^{M-1} P_{ij} + \sum_{i=0}^{N-1}\sum_{j=0}^{M-1} G_{ij}}},
\end{equation}

where $i$ and $j$ represent pixel indices for the height $N$ and width $M$. The value of Dice score ranges between 0 and 1 and a higher score corresponds to a better match between the predicted image $P$ and the ground truth image $G$.

The loss function for DIU-Net is given by \cite{cahall2019inception}:

\begin{equation} \label{eq:3}
\mathcal{L}_{Dice}(P, G) = -\log\left[\frac{1}{K}\sum_{i=0}^{K-1} Dice(P_{i}, G_{i})\right].
\end{equation}

\subsection{Training, Testing, and Evaluation of DIU-Net}
We compared the performance of the proposed DIU-Net with Inception U-Net that did not incorporate dilated modules \cite{cahall2019inception}. We trained both models under same conditions to ensure a fair comparison. Both models were trained using \textit{k}-fold cross-validation scheme with $k=10$. The dataset was randomly split into 10 mutually exclusive subsets of equal or near equal size. Each algorithm was run 10 times subsequently, each time taking one of the ten splits as the validation set and the rest as the training set. In our experiments, each model was trained 10 times using a different set of 90\% of the data and validated on the remaining 10\% data. This resulted in a total of 20 models, i.e., 10 models for U-Net with Inception modules, 10 models for DIU-Net. The Dice scores presented in the Results section are median values of the ten trained models.

We used stochastic gradient descent with an adaptive moment estimator (Adam) for training all models \cite{kingma2014adam}. The initial learning rate was set to $10^{-4}$ which was exponentially decayed every 10 epochs. The batch size was set to 64 and each model was trained for 100 epochs. All learnable parameters, i.e., weights and biases of the models were initialized based on He initialization method. We used Keras application programming interface (API) with TensorFlow backend for the implementation of all models. All models were trained on a Google Cloud Compute instance with 4 NVIDIA TESLA P100 graphical processing units (GPUs). 

After training, each model was tested on the entire BRATS 2018 dataset. For each image, the intra-tumoral structures were combined to produce glioma sub-regions, and Dice scores were computed. The process was repeated for each image, and after evaluating all images, a median Dice score was calculated for each glioma sub-region. Overall, this process generates 2 sets of 10 Dice scores for each glioma sub-region. Each set was then evaluated for normality using the Shapiro-Wilk test, with the probability of Type-I error set to $\alpha  = 0.05 $. Based on the results of the Shapiro-Wilk test, we found that the set of Dice scores were not normally distributed. Therefore, we used non-parametric test, i.e., two-sided Wilcoxon signed rank test to compare Dice scores of two models. 

\begin{figure}[tpb]
\centering
\includegraphics[width=3.3in]{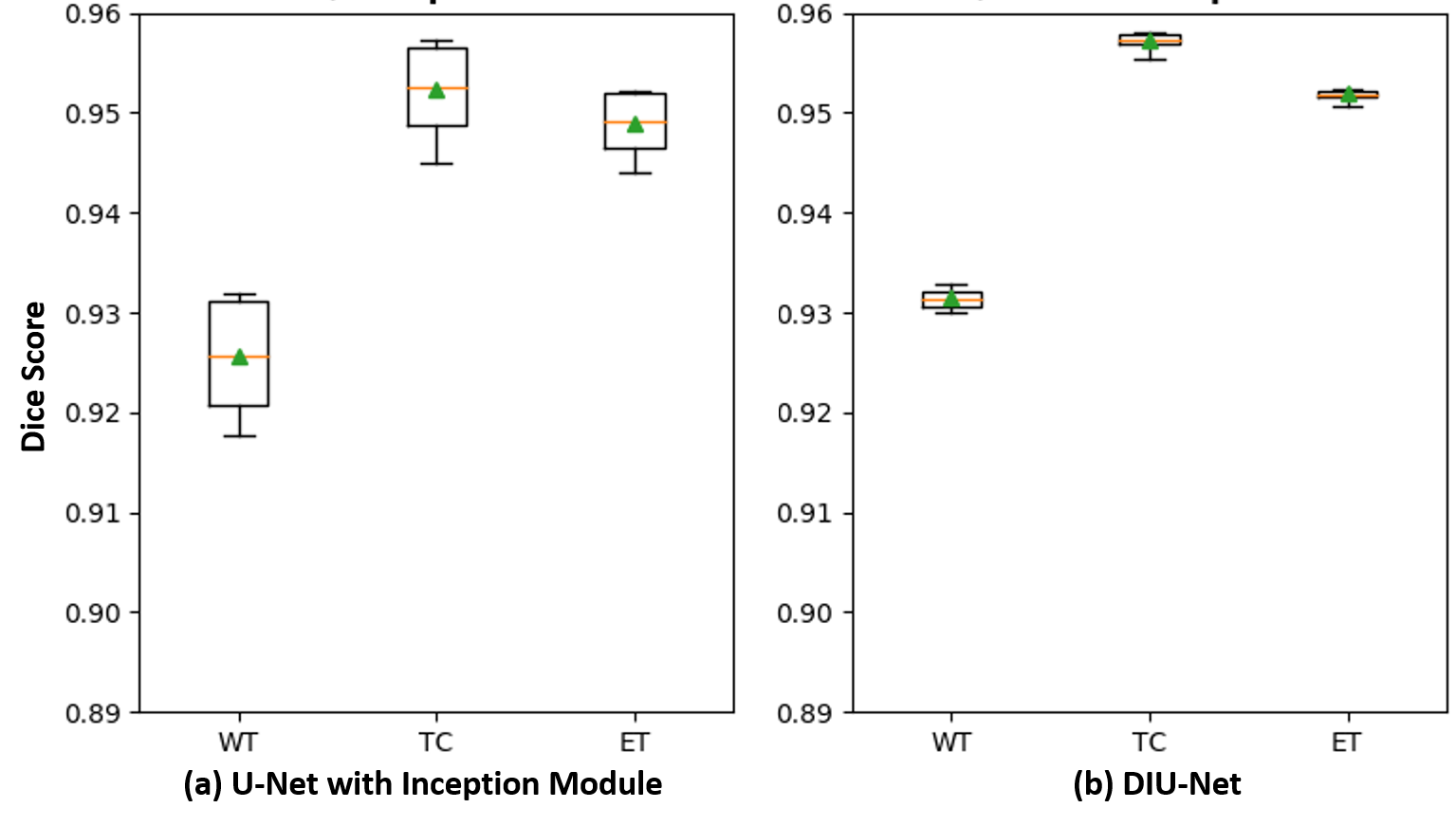}
\caption{\label{fig:dsc_boxplot} Dice scores are presented for both U-Net with Inception module and DIU-Net using box plot. On the x-axis, we present glioma sub-regions including whole tumor (WT), tumor core (TC), and enhancing tumor (ET). The median values are denoted by the horizontal orange line and the mean values are denoted by the green triangle. The increased dice score is statistically significant for WT and TC ($p<0.05$). We also note a significant reduction in the variability for the DIU-Net.}
\end{figure}

\begin{figure}[htpb]
\centering
\includegraphics[width=3.3in]{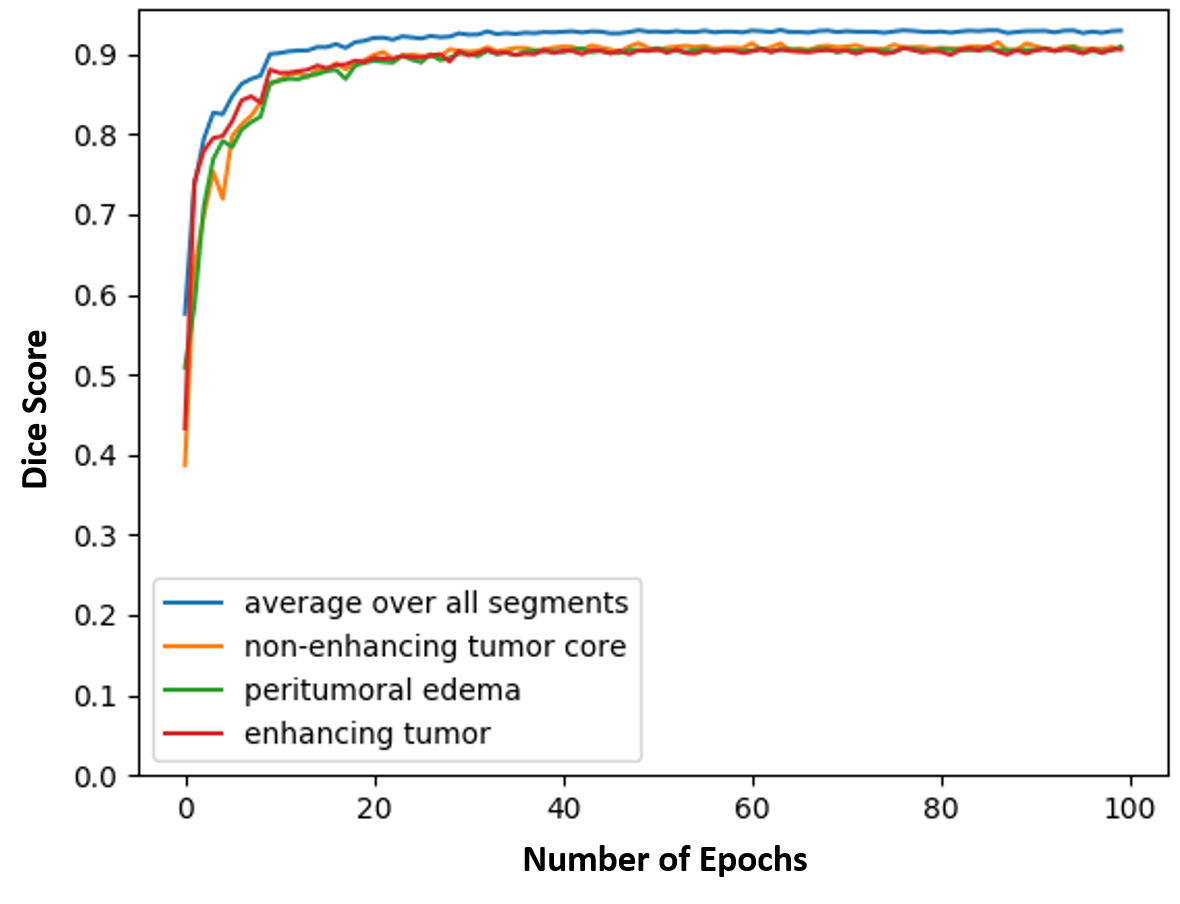}
\caption{\label{fig:val_loss_plot}  Dice scores are presented for an increasing number of epochs separately for each intra-tumoral structure during validation.}
\end{figure}

\section{Results}
We present cross-validation Dice scores for all three glioma sub-regions using the box plot for both models in Fig. \ref{fig:dsc_boxplot}. 
We note that DIU-Net showed significant improvement in the whole tumor sub-region, i.e., Dice score increased from 0.925 to 0.931 with $p < 0.05 $. Similarly, for the tumor core sub-region, the Dice score improved from 0.952 to 0.957 with $p < 0.05 $. However, for the enhancing tumor, the change was not statistically significant, $p = 0.114 $.

\begin{figure*}[t]
\centering
\includegraphics[width=7in]{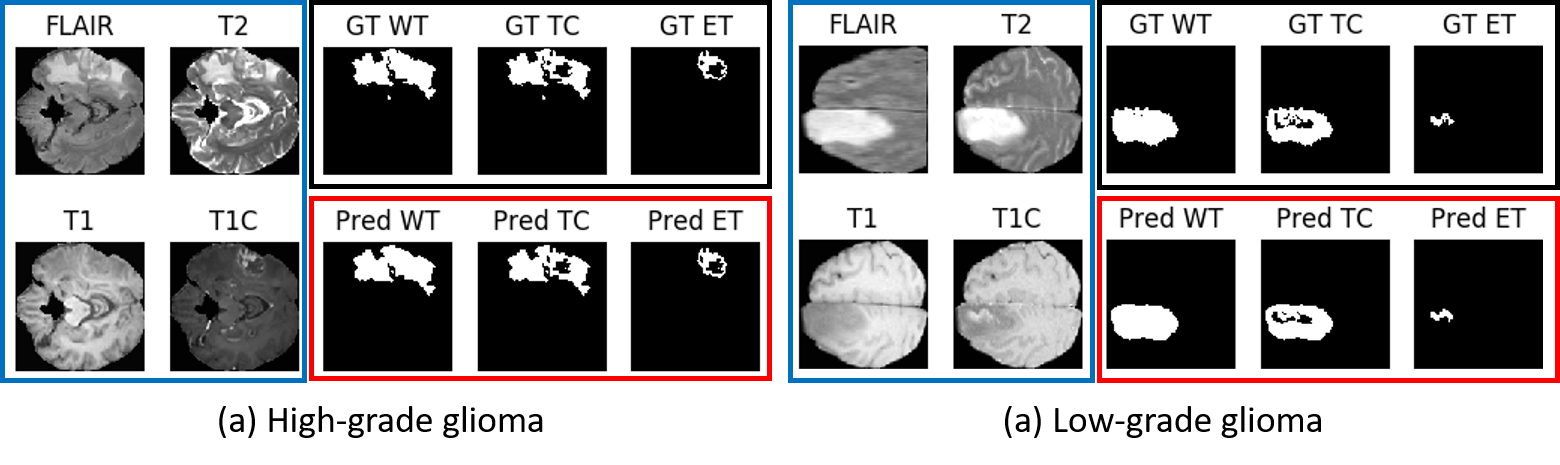}
\caption{\label{fig:lgg} The segmentation results for a representative high-grade, and a low-grade glioma patient are presented. Images in blue blocks are the four MRI modalities. Images in the black blocks (top row) are the ground truth segments and are denoted by ``GT'' for all three glioma sub-regions (whole tumor - WT, tumor core - TC, and enhancing tumor - ET). Images in the red blocks (bottom row) are segmentation results and are denoted by ``Pred''.}
\end{figure*}

The validation Dice score curves plotted against the number of epochs for all intra-tumoral structures for the DIU-Net are presented in Fig. \ref{fig:val_loss_plot}. The improvement in the segmentation over the number of epochs is evident. The segmentation results from one representative high-grade glioma and one low-grade glioma case are presented in Fig. \ref{fig:lgg}. We note that the predicted segments (shown in the red block) of the glioma sub-regions are visually similar to the ground truth segments (shown in the black block). 

\section{Discussion and Conclusions}
We aimed to tackle the challenging problem of pixel-level segmentation in brain MRIs for tumor delineation, which, in turn, is essential for tumor diagnosis, identification, and surveillance. We introduced dilated convolutions in Inception modules and incorporated these modules into the U-Net architecture (DIU-Net). We extended our previously proposed framework and significantly improved its accuracy (measured using the Dice score) \cite{cahall2019inception}. We used \textit{k}-fold cross-validation and found that DIU-Net significantly improved ($p<0.05$) the tumor segmentation performance in two of the three glioma sub-regions, i.e., whole tumor and tumor core.

We hypothesize that there is more contextual information for the whole tumor and tumor core, which DIU-Net was able to capture in the learning process. The results of enhancing tumor suggest that larger contextual information does not benefit model performance for this sub-region. This may be potentially linked to the small number of pixels in this sub-region relative to the other glioma sub-regions. It is essential to mention that DIU-Net is computationally more efficient, i.e., DIU-Net has 2.5 million fewer parameters than the U-Net with Inception modules. DIU-Net achieves significantly better results at a lesser computational cost (15\% fewer parameters). The dice scores for each glioma sub-region are comparable or exceed the results of other recently published architectures, including No New-Net, which achieved second place in the BRATS 2018 competition \cite{isensee2018no}, SDResU-Net \cite{sdresunet}, and the ensemble approach proposed in \cite{kao2018brain}.

\footnotesize
\bibliographystyle{IEEEbibSPMB}
\bibliography{IEEEabrv,IEEESPMB}

\end{document}